\documentclass[%
 reprint,
 amsmath,amssymb,
 aps,
superscriptaddress]{revtex4-2}

\usepackage{graphicx}% Include figure files
\usepackage{dcolumn}% Align table columns on decimal point
\usepackage{bm}% bold math
\usepackage[english]{babel}
\begin{document}

\preprint{APS/RC}

\title{Optical phase encoding in pulsed approach to reservoir computing}

\author{Johan Henaff}
 \affiliation{Laboratoire Kastler Brossel, Sorbonne Universit\'e, ENS-Universit\'e PSL, CNRS, Coll\`ege de France, 4 place Jussieu, 75252 Paris, France}
 
\author{Matthieu Ansquer}
 \affiliation{Laboratoire Kastler Brossel, Sorbonne Universit\'e, ENS-Universit\'e PSL, CNRS, Coll\`ege de France, 4 place Jussieu, 75252 Paris, France}
 \affiliation{Phasics SA, Batiment Mercury I, Espace technologique, Route de l'Orme des Merisiers, 91190 Saint-Aubin, France}

\author{Miguel C Soriano}
 \affiliation{
 Instituto de F\'isica Interdisciplinar y Sistemas Complejos IFISC (CSIC-UIB),  Campus Universitat Illes Balears,  E-07122 Palma de Mallorca, Spain
}
\author{Roberta Zambrini}
\affiliation{
 Instituto de F\'isica Interdisciplinar y Sistemas Complejos IFISC (CSIC-UIB), Campus Universitat Illes Balears,  E-07122 Palma de Mallorca, Spain  
}

\author{Nicolas Treps}
\affiliation{Laboratoire Kastler Brossel, Sorbonne Universit\'e, ENS-Universit\'e PSL, CNRS, Coll\`ege de France, 4 place Jussieu, 75252 Paris, France}

\author{Valentina Parigi}
\affiliation{Laboratoire Kastler Brossel, Sorbonne Universit\'e, ENS-Universit\'e PSL, CNRS, Coll\`ege de France, 4 place Jussieu, 75252 Paris, France}

\date{\today}

\begin{abstract}
The exploitation of the full structure of multimode light fields enables compelling capabilities in many fields including classical and quantum information science. We exploit data-encoding on the optical phase of the pulses of a femtosecond laser source for a photonic implementation of a reservoir computing protocol. Rather than intensity detection, data-reading is done via homodyne detection that accesses combinations of amplitude and phase of the field. Numerical and experimental results on NARMA tasks and laser dynamic predictions are shown. We discuss perspectives for quantum enhanced protocols.
\end{abstract}

\maketitle

Reservoir computing is a supervised machine learning approach to time series processing, rooted in recurrent neural networks \cite{lukosevicius_reservoir_2009,book_RC}. Inspired by the brain mechanisms, many interconnected artificial neurons process input information and display an internal memory. Recurrent neural networks are then suitable for temporal tasks such as speech recognition \cite{verstraeten2006reservoir, larger_high-speed_2017}, but at the expense of being hard to train. All the weights of the networks need indeed to be trained using backpropagation through time \cite{werbos1990backpropagation}, a time-consuming and not always converging, procedure \cite{doya1992bifurcations}.
Differently, in the reservoir computing (RC)  approach only the weights of the output layer are trained to process information \cite{jaeger_harnessing_2004, jaeger_echo_nodate}. Those architectures are composed of three elements: an input layer to inject the data into the system, a reservoir composed of a large number of neurons (or nodes) randomly connected, and an output (or reading) layer to extract the information from the reservoir.  
Under certain conditions on the reservoir, training the output layer with a simple linear regression can be sufficient \cite{lukosevicius_reservoir_2009,jaeger_echo_nodate}.

In this paper, we present the design of a reservoir computing protocol using a single non linear node with delayed feedback as in \cite{appeltant_information_2011}. 
While recent works have successfully implemented reservoir and neuromorphic computing via the frequency components of optical frequency combs  \cite{Xu23,Butschek22,lupo2023deep}, here we exploit the time features, i.e. the pulses basis, of an optical frequency comb as nodes of the reservoir. Moreover coherent homodyne detection is used, so that information can be encoded in the phase components of the field rather than its intensity or amplitude. We show that despite a small number of nodes and a low non-linearity, our protocol has good performances, displaying both nonlinear memory and forecasting capabilities. Our system builds on the notion that optical pulses can be used to build spiking reservoirs \cite{owen2022ghz,owen2023photonic} and that phase encoding for information injection can yield to better performance in photonic reservoir computers \cite{Bauwens22,kanno2022reservoir}.

Optical based computing \cite{mcmahon23} might be able to give speed or energy-efficiency benefits over electronics. Among the peculiar features of optical systems for computing, time and frequency multiplexing can be exploited in our system. 
In addition, the large signal-to-noise ratio characteristics of the detection pave the way for quantum protocols  \cite{Mujal21} by exploring quantum features of the optical states. Recently demonstrated frequency and time-multiplexed quantum platforms  \cite{Koadou23,Roman23}  can then be exploited in promising protocols \cite{nokkala2021gaussian,GarciaBeni23}.

\begin{figure*}[t]
\centering
\includegraphics[width=1\textwidth]{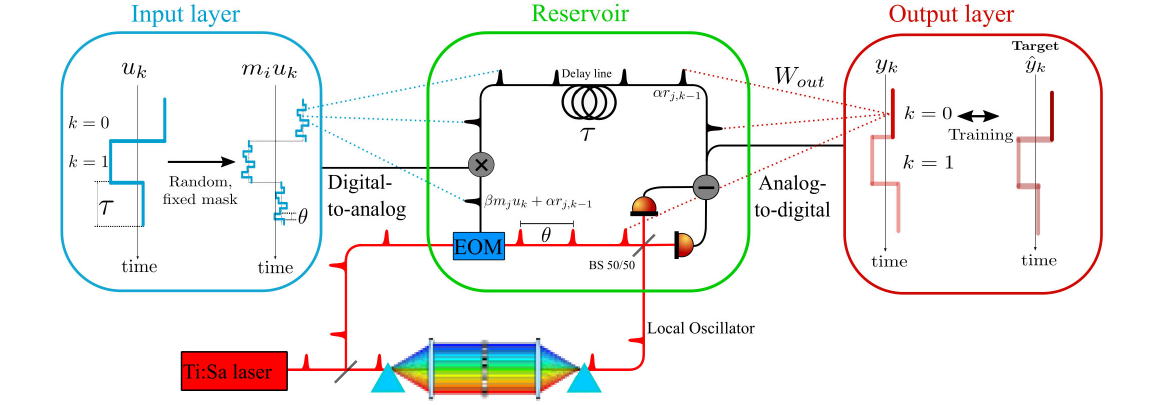}
\caption{\textbf{Experimental scheme of the delay-based reservoir computing architecture}. The optical field generated by a Ti:Sa laser is separated in two. The weak beam, the signal, is sent into an electro-optic modulator (EOM) controlled by an arbitrary waveform generator (AWG). Information is encoded in the phase of each pulse by applying a voltage to the EOM. The strong beam, the local oscillator, is sent into a pulse shaper to match the dispersion introduced by the EOM on the other arm. Both beams are combined on a 50-50 beam-splitter and detected by a high bandwidth balanced detector. After low pass filtering, to control the parameter $T_{BW}$, the signal is sent into a long coaxial cable to create a delay. A part of this signal is then mixed with the input data to create the feedback. The other part of the signal is acquired to proceed to the off-line training of the output weights. Optical and electronic pulses are represented in red and black, respectively.}
\label{fig:manip_RC}
\end{figure*}

In a delayed feedback RC architecture, the reservoir is composed of virtual neurons distributed in time in a feedback loop \cite{appeltant_information_2011}  and only one neuron is accessible at a time. Each neuron is separated from the others by a time interval $\theta$. In the following, we consider $V$ neurons  labeled by $j$. The time-span of the reservoir, $\tau$, is equivalent to its size. It is given by $\tau = V \theta$.\\
To process a time sequence $u(t)$, the signal is sampled according to $u_k = u(t = k T_s)$, where $T_s$ is the sampling time and $k \in \left[ 1 : L \right]$, with $L$ the length of the input sequence. Each data point $u_k$ is distributed on every node in the time-span of the reservoir. As illustrated in Fig. \ref{fig:manip_RC} (input layer), this is achieved by applying a mask function $m(t_j)$ to $u_{k}$. This mask is composed of $V$ individual steps of duration $\theta$ so that $m_j = m(t_j = j \theta)$ \cite{brunner_5_2019}.\\
The state of neuron $r_{j, k}$, corresponding to the $j^{th}$ node for the data point $k$, at time $t_j = k \tau + j \theta$, is given by a non-linear function of the sum of input data and the previous  state of the reservoir $r_{j,k-1}$. The specific description of our reservoir system is detailed later in Eq. \ref{eq:MandPhi}.
\\
The state of the reservoir is extracted at the output layer $y_k$, given by a simple linear combination of the reservoir neuron values: 
\begin{equation}
    y_k = \mathbf{W}_{out}^T \mathbf{r}_k,
\end{equation}
where $\mathbf{r}_k$ is the vector of $V$ neurons (measured signals), $\mathbf{W}_{out}$ is the vector containing the output weights, and $T$ stands for the transposition of the vector column. \\
To perform the supervised training, a set of data of length $L_{train}$ is taken. The reservoir states $\mathbf{r}_k$ and the target data $\hat{y}_k$ for the full set of training data are regrouped in the vectors $R =( \mathbf{r}_1, \cdots \, \mathbf{r}_{L_{train}})$ and $\mathbf{\hat{Y}}^T= (\hat{y}_1, \cdots, \hat{y}_{L_{train}})$. 
The training procedure is then reduced to the inversion of a matrix so that $\mathbf{W}^{\star}_{out}$, the trained set of output weights, is given by $\mathbf{W}^{\star}_{out} = \mathbf{\hat{Y}} R^{-1}$. Note that in practice more complex transformations are used, such as Ridge regression \cite{lukosevicius_reservoir_2009}, to prevent the protocol from overfitting. The accuracy of the training is assessed on a new set of data, called test-set, of length $L_{test}$. It is given by computing an error function such as the normalized root mean squared error, or a correlation coefficient such as Pearson coefficient.

As previously mentioned, in this work, the reservoir is composed of a succession of laser pulses. The information is encoded in the phase of each pulse and extracted using homodyne detection. In the following, we derive the expressions corresponding to the experimental configuration presented in Fig. \ref{fig:manip_RC}.

To describe the experiment, let us start with the expression of the electric field of one pulse given by $E^{(+)}_{\mathrm{pulse}}(t) = \mathcal{E}_0 a(t) e^{-i \omega_0 t}$. Even though a high bandwidth detector is used to measure the field pulse by pulse, this detector is not fast enough to resolve the pulse shape. In addition, one data point is taken for each pulse. Thus, we consider that each pulse has some average amplitude $a$. The information, i.e the input data, can be encoded in the two degrees of freedom of the pulses, their amplitude or their phase. In \cite{nokkala2021gaussian}, Nokkala et. al. investigated several RC protocol architectures with different encoding for the input data. In particular, encodings into the amplitude or the phase of a coherent state were compared. They demonstrated numerically that phase encoding can lead to better performance. Therefore, using the fact that under certain conditions the laser field can be assimilated to a coherent state, we choose to encode the information in the phase of the pulses. Hence, the field of the $j^{th}$ pulse at the time step $k$ is multiplied by a phase term containing the injected data at step $k$ and the measurement done at step $k-1$ according to 
\begin{equation}\label{eq:pulse_rc_node}
E^{(+)}_{j,k} = \mathcal{E}_0 a \exp \left[ i(\beta m_j u_k + \alpha M_{k-1,j}) \right] , 
\end{equation}
where $E^{(+)}_{j,k}$ is the complex field amplitude of the pulse at time $t_n = k \tau + j \theta$.  As shown in the experimental scheme in Fig. \ref{fig:manip_RC},  data  $m_j u_k $ are summed to homodyne measurements    $M_{k-1,j}$ at step $k-1$  via a mixer, in order to provide memory to the system. 
The gain parameters $\alpha$ and $\beta$ are set by electronic transfer functions and  can be used to optimize the protocol. 
The homodyne detection measurement $M_{k,j}$ of the node $j$ at time step $k$, providing the necessary non-linearity for the protocol to work, is then given by:
\begin{eqnarray}
M_{k,j}&=&C \sin(\beta m_j u_k + \alpha M_{k-1,j}),
\label{eq:model_continuous_desynch}
\end{eqnarray}
where $C$ is a constant depending on the detector characteristics and the experimental conditions.

In order to create links between the virtual nodes of the reservoir, we use the finite bandwidth of the system. As it will be detailed later, the homodyne detector has a response time that is a bit slower than the pulse repetition rate corresponding to the node separation ($\theta = T_r \simeq 6.4$ ns). The bandwidth of the detection determines the parameter $T_{BW}$ of the protocol. In addition, low-pass filters can be added after the detector to reduce even more the effective bandwidth $T_{BW}$ and tune the parameter $T_r/T_{BW}$. This ratio determines the corresponding graph connectivity of the reservoir \cite{brunner2018tutorial}. Consequently, the measurement of the node $j$ at time step $k$ is given by
\begin{eqnarray}\label{eq:MandPhi}
    \begin{cases}
      M_{k,j}\simeq C \sin(\phi_{k,j-1})e^{-T_r/T_{BW}}\\
      \hspace{1.2cm}+ C \sin(\phi_{k,j})(1-e^{-T_r/T_{BW}})\\
      \phi_{k,j}=\beta m_j u_k + \alpha M_{k-1,j}
      \end{cases}
\end{eqnarray}
From equation (\ref{eq:MandPhi}), we see that each node is connected to a few previous nodes with different strengths. Those connections depend on the ratio $T_r/T_{BW}$. Changing this ratio modifies the influence of a given node on the future ones.

The scheme of the experimental setup is further illustrated in Fig. \ref{fig:manip_RC}. As usual in homodyne detection, the field from the laser is split into two, a weak beam, the signal, and a strong beam, the local oscillator (LO). The phase encoding is achieved using an electro-optic modulator (EOM) on the signal arm. By applying a voltage to this modulator, the phase of each pulse can be modified to encode the information. This EOM is driven from a computer via  an arbitrary waveform generator (AWG). For simplicity, we decided to implement the feedback electronically as in \cite{appeltant_information_2011} and not optically as in \cite{brunner_parallel_2013}. An optical feedback would require an optical loop. Such loops are usually built using fiber elements, which are not easy to handle with femtosecond pulses, mainly due to chromatic dispersion. As long delays are difficult to handle in free space, we have chosen the electronic feedback. This is achieved by first detecting the electric field after the encoding via an homodyne detection, which is equivalent to measuring the nodes of the reservoir. The detection is performed by a home-made high-bandwidth balanced detector. Its bandwidth is approximately 100 MHz and details about this detector can be found in \cite{kouadou2021}. To create the links between the nodes, a low pass filter is introduced after the detection ($T_{BW}\simeq21ns$). 
After detection, the measured signal is split in two. A small part is gathered for off-line processing and the other part is sent in a long coaxial cable to create the feedback loop that provides the reservoir memory. The delayed signal is finally mixed with the input signal from the AWG to be sent to the EOM, creating the feedback loop. At the time of the experiment, the longest cable in the lab with sufficiently low losses could only handle a delay corresponding to V=35 nodes thus defining the size and corresponding maximum memory available to the RC system.

To characterize the computing abilities of the setup, we first report the results for an academic benchmark,  namely the NARMA (Nonlinear AutoRegressive Moving Average) task. It involves predicting the next output value in a sequence based on a non-linear combination of past output values and input data. It is defined by \cite{atiya2000new,jaeger2002adaptive}
\begin{eqnarray}
    y(t) &=& 0.3y(t-1) + 0.05y(t-1)\sum_{i=t-N}^{t}y(i) \nonumber \\
    &+& 1.5u(t-1)u(t-N) + 0.1,
\end{eqnarray}
where the input $u$ is taken from a uniform random distribution in the interval $[0,0.5]$. The  parameter  $N$ determines how many time steps in the past of inputs in $u$ influence the current output $y$. The drawn sequence $u$ is used as the input to our reservoir, which is then trained on the corresponding target output $y$. The performance of the protocol is evaluated by the Pearson correlation coefficient between the NARMA-$N$ value $y$ and the predicted value $\bar{y}$.\\

\begin{figure}[h]
    \centering
    \includegraphics[width=0.49\textwidth]{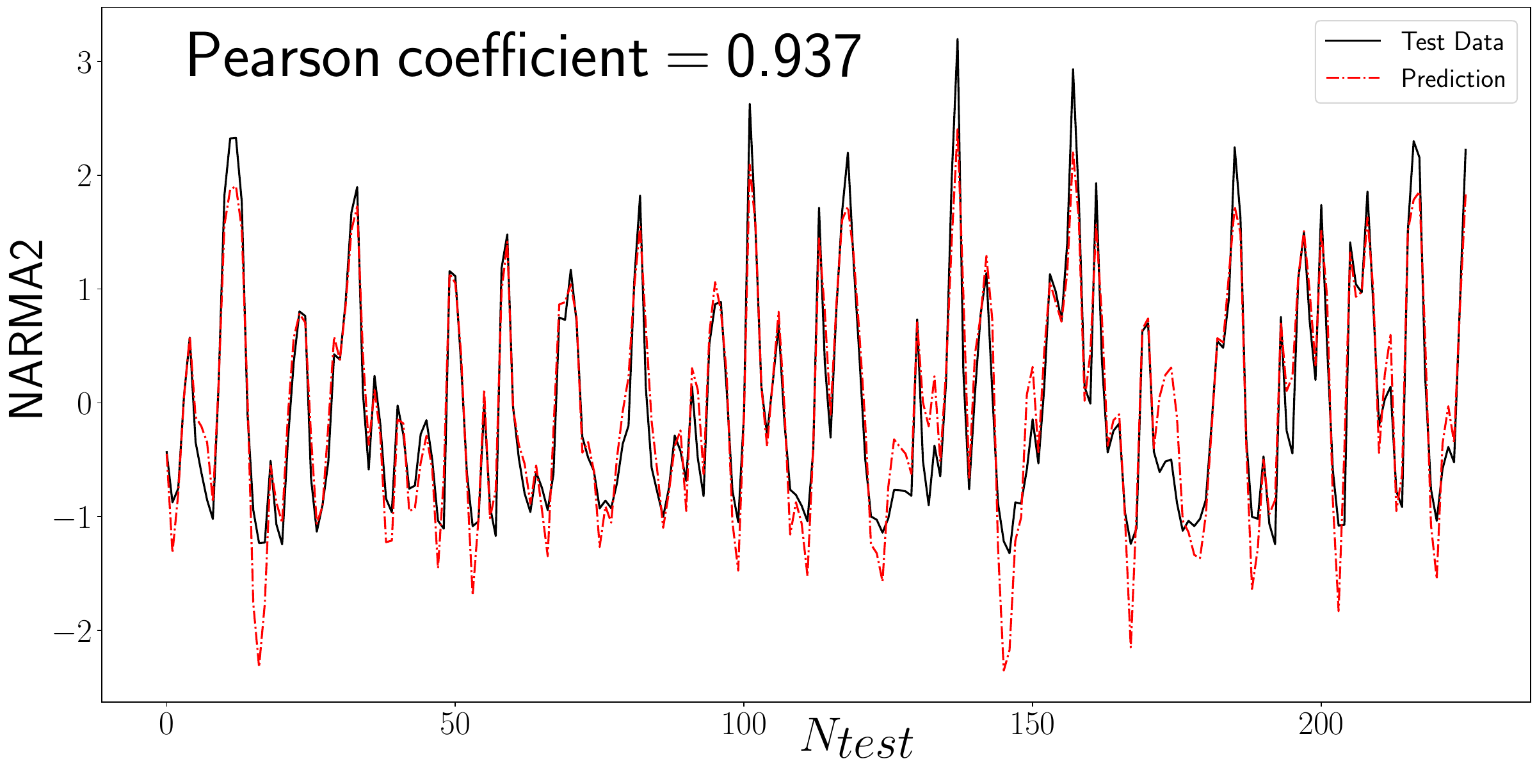}
    \caption{\textbf{Experimental prediction of NARMA-2 function.} The protocol was trained on 2250 input data and tested on 600.}
    \label{fig:NARMA2}
\end{figure}

\begin{figure}[h]
\centering
\includegraphics[width=0.5\textwidth]{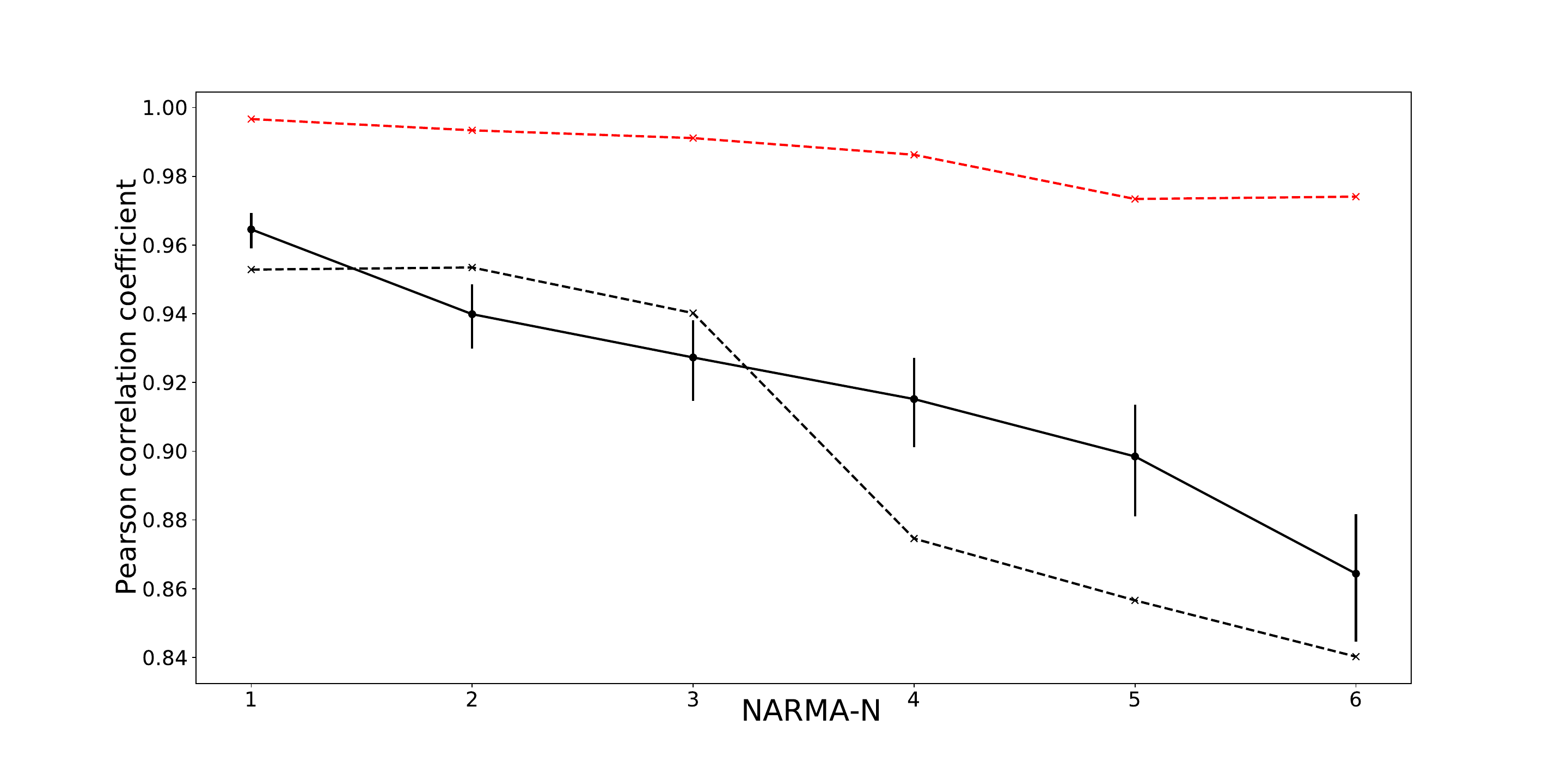}
\caption{\textbf{Pearson correlation coefficient as a function of the input delay parameter $N$.} Experiment (solid line) and simulation (dashed line) were trained on 2250 input data and tested on 600. Black lines correspond to the experimental and numerical results for 35 virtual nodes. Increasing the number of nodes to 100 (red) reduces the error. The simulations have been performed for $\alpha=0.7$ and $\beta=1.0$.}
\label{fig:ErrorVNodes}
\end{figure}

As shown in Fig. \ref{fig:NARMA2}, for a small value of input delay parameter ($N=2$) the experimental RC protocol can approximate the target data. The accuracy of the approximation decreases monotonically for higher values of $N$ as shown in Fig. \ref{fig:ErrorVNodes} (solid line). 
In addition to the optical experimental protocol, we  performed  numerical simulations. They are run according to  Eq. (\ref{eq:MandPhi}),  given the known characteristics of the experiment (e.g. detection noise, sampling rate in acquisition). 
We find that the performances of the experimental setup are qualitatively similar to the simulated ones, see dashed black line in Fig. \ref{fig:ErrorVNodes}. The discrepancies between experiment and simulation may be due to the presence of other experimental noise sources, such as phase drifts, not accounted for in the numerical simulations. 
It is also worth mentioning that the $\alpha$ and $\beta$ parameters can be finely tuned in the simulation while in the experiment the $\alpha$ parameter was set with an electronic attenuator that didn't allow fine tuning, therefore we  couldn't reach identical $(\alpha,\beta)$.
Even though a quantitative agreement is not possible, the numerical simulations allows us to estimate the impact of an increasing reservoir size. 
The red line in Fig. \ref{fig:ErrorVNodes} indicates that the error in the NARMA-$N$ task is greatly reduced when the number of nodes is increased to 100. 
Similar trends can be expected to happen if the reservoir size could be increased in the experimental implementation.

To give an example of a practical application of this machine learning protocol, following the idea of \cite{amil_machine_2019, cunillera_cross-predicting_2019}, we aim to predict the dynamics of the fluctuations of the laser studied in \cite{FcombDynamicsAnsquer2021}. It was demonstrated that the amplitude and phase noises are mainly induced by the pump laser intensity noise. Consequently, in this section, we try to predict the power $\delta\epsilon$ and central frequency $\delta\omega_c$ fluctuations from the intensity noise of the pump laser. 

\paragraph*{}
As we aim to predict the intensity related dynamics, the intensity fluctuations of the pump power are used as input data. They are measured with a photodiode from a leak in the laser cavity. The output data, used as target, are the fluctuations of each laser variable, measured using the experimental scheme described in \cite{FcombDynamicsAnsquer2021}. Those quantities are recorded  simultaneously. Each acquisition is composed of 10000 data points. We take 8000 data points to train the reservoir. The 2000 remaining points are used as a validation set to see how well the algorithm performs. 
\paragraph*{}   

\begin{figure}[h]
\centering
\includegraphics[width=0.48\textwidth]{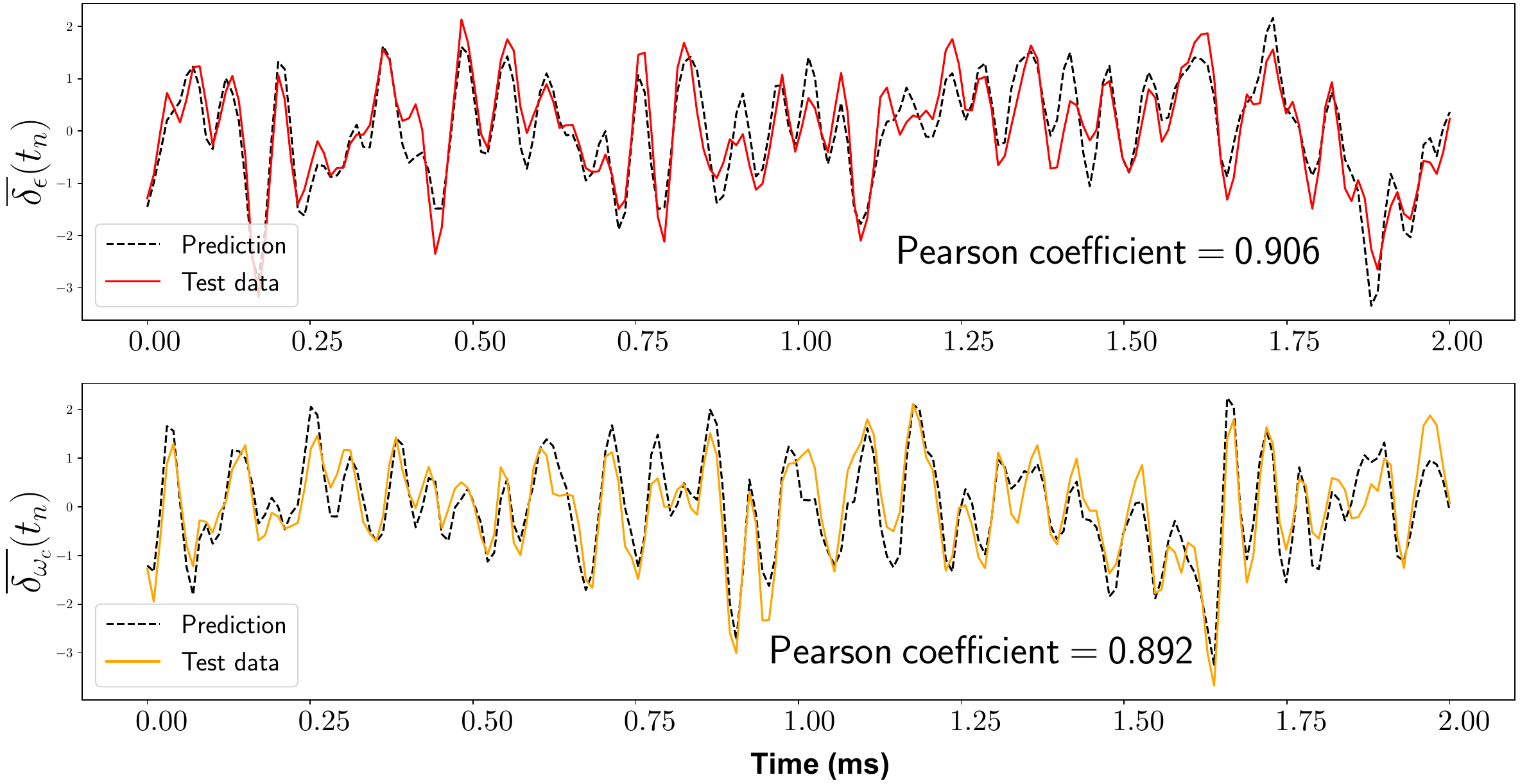}
\caption{\textbf{Performance of the protocol on laser dynamics prediction.} The input data used for the training are the fluctuations of the pump laser power. The laser power fluctuation is in red and the central frequency fluctuation in yellow.}
\label{fig:exp_noise_prediction}
\end{figure}

The mean power noise and the center spectrum, predicted after training from the pump laser fluctuations together with the expected signal are presented on Fig. \ref{fig:exp_noise_prediction}. It can be seen that the experimental prediction is close to the real measured noise in both cases. Given these positive results, one could think of an active stabilization of the laser using the reservoir computing approach combined with a feedback control loop.

In conclusion we implemented an optical reservoir computing protocol exploiting the pulse basis of an optical frequency comb. The use of a homodyne detection for readout allowed us to encode the information in the phase of the coherent states. This experimental proof-of concept has a relatively small number of virtual nodes and a single hardware node that operates in a low nonlinearity regime, yet is proven to perform non-trivial computational tasks. This work paves the way toward more refined protocol with an all-optical configuration. 
\\
Moreover, recent work in the squeezing generation at high repetition rate \cite{Koadou23} could be used to implement a quantum reservoir computing protocol with a scheme similar to the one of  this article. The use of quantum states and the spectral multimode nature of our frequency comb could be harnessed to improve the computational capacity of our protocol as suggested in \cite{GarciaBeni23}.

\subsection*{Funding} European Research Council under the Consolidator Grant COQCOoN (Grant No. 820079) and the Agence Nationale de la Recherche with support of  the Direction G\'en\'erale de l'Armement (LASAGNE ANR-16-ASTR-0010-03).  We acknowledge the Spanish State Research Agency, through the Mar\'ia de Maeztu project CEX2021-001164-M funded by the MCIN/AEI/10.13039/501100011033 and through the COQUSY project PID2022-140506NB-C21 funded by MCIN/AEI/10.13039/501100011033, MINECO through the QUANTUM SPAIN project, and EU through the RTRP - NextGenerationEU within the framework of the Digital Spain 2025 Agenda.

\subsection*{Disclosures} The authors declare no conflicts of interest.

\subsection*{Data availability} Data underlying the results presented in this paper are not publicly available at this time but may be obtained from the authors upon reasonable request.

\nocite{*}

\bibliography{apssamp}

\end{document}